\documentclass[twocolumn,showpacs,preprintnumbers,amsmath,amssymb]{revtex4}

\usepackage{graphicx}
\usepackage{dcolumn}
\usepackage{bm}

\begin{document}


\title{A brief review of a modified relativity that explains cosmological constant}  
\author{Cl\'audio Nassif Cruz}
\altaffiliation{{\bf UFOP}: Universidade Federal de Ouro Preto, Morro do Cruzeiro, Bauxita, 35.400-000-Ouro Preto-MG, Brazil. \\
email: claudionassif@yahoo.com.br}
\author{A. C. Amaro de Faria Jr}
\altaffiliation{{\bf UTFPR-GP}: Federal Technological University of 
Paran\'a, Av. L. Bastos, 85053-525 , Guarapuava-PR, Brazil. 
\\
email: atoni.carlos@gmail.com} 



\date{\today}

\begin{abstract}
The present review aims to show that a modified space-time with an invariant minimum speed provides a relation  with Weyl geometry in the Newtonian approximation of weak-field. The deformed Special Relativity so-called 
Symmetrical Special Relativity (SSR) has an invariant minimum speed $V$, which is associated with a 
preferred reference frame $S_V$ for representing the vacuum energy, thus leading to the cosmological constant
($\Lambda$). The equation of state (EOS) of vacuum energy for $\Lambda$, i.e., $\rho_{\Lambda}=\epsilon=-p$ emerges 
naturally from such space-time, where $p$ is the 
pressure and $\rho_{\Lambda}=\epsilon$ is the vacuum energy density. With the aim of establishing a relationship between $V$ and $\Lambda$ in the modified metric of 
the space-time, we should consider a dark spherical universe with Hubble radius $R_H$, having a very low value 
of $\epsilon$ that governs the accelerated expansion of  universe. In doing this, we aim to show that SSR-metric has an equivalence with a de-Sitter (dS)-metric 
($\Lambda>0$). On the other hand, according to the Boomerang experiment that reveals a slightly accelerated expansion of the universe, SSR leads to a dS-metric with an approximation for $\Lambda<<1$ close to a flat space-time, which is in the $\Lambda CDM$ scenario where the space is quasi-flat, so that $\Omega_{m}+\Omega_{\Lambda}\approx 1$. We have $\Omega{cdm}\approx 23\%$ by representing dark cold matter, $\Omega_m\approx 27\%$ for matter and $\Omega_{\Lambda}\approx 73\%$ for the vacuum energy. Thus, the theory is adjusted for the redshift $z=1$. This corresponds to the time $\tau_0$ of 
transition between gravity and anti-gravity, leading to 
a slight acceleration of expansion related to a tiny value of $\Lambda$, i.e., we find $\Lambda_0=1.934\times 10^{-35}s^{-2}$. This result is in agreement with observations. 
\end{abstract} 

\maketitle

\section{\label{sec:level1} Introduction} 

Einstein\cite{Einstein} introduced in his paper 
on Special Relativity (SR) a fundamental change in laws
of Newtonian mechanics in order to preserve the
covariance of Maxwell equations. This led to the invariance of the speed of light $c$ in vacuum for any inertial motion. Thus $c$ is the maximum limit of speed
in nature. In view of this, the space, time, mass and energy become related between themselves, as all these quantities depend on speed. However, SR was built on an empty space, i.e., there is no kind of {\it aether} or no vacuum energy in SR, as the uncertainty principle (the zero-point energy) does not belong to the space-time of SR. Our great challenge is the natural inclusion of the quantum principles associated with a fundamental vacuum energy into a new structure of space-time, where the cosmological constant emerges naturally from such first principles that should be deeper investigated.  

The advantage of Symmetrical Special Relativity (SSR) as a theory of Modified Relativity is its kinematic basis that is based on new relativistic effects as consequence of a minimum speed that prevents rest, supporting the principle of uncertainty that prevents completely the certainty on the momentum and the infinite uncertainty on position, still because a plane wave with a well-defined momentum is a quantum idealization for extreme cases used for some unrealistic ideal purposes, as for instance the wave function of a plane wave of a non-relativistic particle inside an one-dimensional box. 

We aim to investigate a modified space-time with the presence of a minimum speed $V$, which is a kinematic invariant at lower energies, as is also the speed of light $c$ for higher energies, by forming a fundamental symmetry of motion that should be justified by first principles (see ref.\cite{N2012}), which must be consistent with the quantum principle related to the zero-point energy given by the uncertainty principle\cite{N2012}. 

Such zero-point energy is represented by the vacuum energy that plays the role of the cosmological constant. In this sense, wouldn't it be natural to realize that  rest is not compatible with the zero-point energy, even more because the fundamental zero-point energy has gravitational origin, so that the particle is not completely free of gravity by forming a bound state with the whole universe? If it is so, we are motivated to build a modified relativity as suggested by Nassif\cite{N2012}\cite{3} in order to become compatible with such vacuum energy or zero-point energy, so that we are motivated to postulate a new kinematic invariance for low energies, i.e., an invariant minimum speed $V$ to be better explored and justified later. 

Actually, the minimum speed $V$ must be invariant because there would be no referential that nullify it, otherwise we would be returning to the classical concept of rest, which is not allowed in this quantum space-time. Such invariance of $V$ will be shown later by means of new velocity transformations, where the invariance of $V$ is represented by a preferred (universal) reference frame $S_V$ given by a cosmic background field (vacuum energy) as explanation for the cosmological constant. 

In sum, we will better understand the implication of the cosmological constant within a new kinematic scenario described by a modified relativity with an invariant minimum speed related to a preferred frame given by the vacuum energy\cite{N2012}\cite{3}. 

So, finally we think that the symmetry of motion due to $c$ and $V$ in the space-time of SSR works like a de-Sitter (dS) space-time. In view of such an equivalence between SSR and a dS space-time, we can get $\Lambda$, such that we will find $\Lambda=\Lambda_0=1.934\times 10^{-35}s^{-2}$ for $z=1$ in the zero gravity-limit when the anti-gravity begins to emerge, i.e., the accelerated cosmic expansion comes into play. 
 
The search for understanding the nature of $\Lambda$ has been the issue of hard investigations\cite{1}\cite{2}. 
The relationship between the vacuum energy density 
$\rho_{\Lambda}$ and $\Lambda$ is well-known, namely
$\rho_{\Lambda}=\Lambda c^2/8\pi G$. 

The fine structure constant $\alpha$ is related to 
$\Lambda$\cite{hgn,hgn2,hgn3}. A variation of 
$\alpha$\cite{sergio,sergio1,sergio2,pad,pad2} could show a fundamental change in the subatomic structure, as 
$\alpha$ is able to connect the micro and macro-world, whose age is $R_H=cT_H$, where $T_H(\cong 13.7$ Gyear) is the Hubble time and $R_H(\sim 10^{26}m)$ is the Hubble radius that represents the visible universe.  

There should be a relationship between $\alpha$ and 
$\Lambda$, which represents the dark universe. The dark
universe is represented by models that search for an 
explanation of the anti-gravity emerging from the dark energy, which has been based on scalar fields\cite{cine1,cine2,cine3,cine4,cine5,cine6,cine7}. 

The emergence of an invariant minimum speed $V$ is associated with a preferred reference frame $S_V$. This leads to a new relativity with Lorentz symmetry violation at lower energies close to $V$. Such a new relativity is so-called Symmetrical Special Relativity (SSR) with many 
fundamental implications.\cite{N2016,N2010,Rodrigo1,Rodrigo2,Rodrigo3,N2018,N2019,N2020,uncertainty}, by also including the Gravitational Bose Einstein Condensate (GBEC)\cite{N2019}, working like the central core of a star of gravitational vacuum so-called gravastar\cite{N2019}, where one introduces a cosmological constant (anti-gravity).  

It has also been shown that some aspects of SSR presents a certain analogy with the principle of
Mach\cite{Rodrigo3,mach,mach2,mach3}, but within a quantum scenario, in the sense that vacuum is responsible for the masses of particles instead of the ``fixed'' stars as supported by Mach's principle as being a classical principle. 
 
There is a relationship between $\alpha$ and 
$\Lambda$, i.e., $\Lambda\propto\alpha^{-6}$\cite{hgn,hgn2,hgn3}. It has been explored that $\Lambda$ is also related to other universal constants as the mass of the electron $m_{e}$, Planck constant ($\hbar$) and the constant of gravity $G$. Thus $\Lambda$ is connected to the standard model constants, namely $\Lambda\sim (G^2/\hbar^4)(m_e/\alpha)^6$\cite{hgn,hgn2,hgn3}. 

Section 3 investigates the space-time and velocity transformations for ($1+1$)D in SSR. 

Section 4 shows that the SSR-metric is equivalent to a dS-metric. Thus $\Lambda$ will emerge naturally from SSR in this scenario, where we will build a model of spherical universe with Hubble radius filled by a vacuum energy density $\rho_{\Lambda}=\epsilon$ with $\Lambda\sim 10^{-35}s^{-2}$. 

In Section 5, by making the approximation for a very weak anti-gravity in the dS-metric from SSR, i.e., $\Lambda<<1$, we are
within a more realistic cosmological scenario of a slightly accelerated expanding quasi-flat space-time, according to the observational data provided by the Boomerang experiment. So, we can go even further by obtaining the tiny numerical value of 
$\Lambda=1.934\times 10^{-35}s^{-2}$ given by the observations at the redshift $z=1$ so-called zero-gravity limit\cite{boomerang}\cite{boomerang1}. 

Section 6 is dedicated to the Weyl geometrical structure of SSR. In the Weyl scenario of conformally flat spacetimes, we will show in a simple and direct way that the
factor $\Theta(v)$ in Eq.(3) of SSR behaves like a conformal Weyl factor, so that SSR includes a Weyl conformal geometry in the regime of Newtonian 
weak-field, i.e., for $\phi/c^2<<1$, such that $\Theta\cong 1$, which is the own conformal factor of SSR given for 
the weak-field regime, where the space-time is almost flat. 

The great relevance of this regime of weak-field in the Weyl structure is that such regime corresponds to the slight acceleration of the universe for 
$z=1$, where occurs a transition between gravity and 
anti-gravity. So we will conclude that the current 
expanding universe is governed by a Weyl conformal geometry for weak-field by representing an almost flat space-time as a particular case of Eq.(3). 

It is important to notice that the Weyl structure was originally proposed with the aim of presenting a unification model between Electromagnetism and Gravitation. The purpose and importance of this work is to show that the Weyl structure emerges from SSR at the weak field boundary. Specifically speaking, the Weyl field is represented by the conformal structure of the theory related to a quasi-flat space-time metric. In this sense, we show that the SSR conformal factor 
$\Theta$\cite{Rodrigo1}, in addition to being conformally flat, is directly related to the Weyl factor with the same approximation in the weak field limit,
i.e, $\Theta\approx 1$ ($\phi/c^2<<1$). This result is relevant, since it shows that the Minkowski space-time metric is slightly perturbed by showing that the Weyl structure, to some extent, manifests itself in the weak-field boundary of SSR. 

In the cosmological scenario, such weak-field regime ($\Theta\approx 1$) occurs for supernovaes of type $1A$ with redshift $z\approx1$ when occurred the transition from gravity to anti-gravity with a slight acceleration represented by a very small positive cosmological constant to be obtained according to the observational data of the Boomerang experiment.

Finally, in Section 7 we investigate the cosmological implications of the isotropic background field (vacuum energy) related to the invariant minimum speed $V$. This leads to a negative pressure at the cosmological scales (cosmological anti-gravity), which is represented by the equation of state (EOS) of vacuum for the cosmological constant, i.e., 
$\rho_{\Lambda}=\epsilon=-p$. 

\section{\label{sec:level1} An invariant minimum speed in the space-time} 

The motivations for considering the existence of a minimum speed for very low energies ($v<<c$) are given 
by the following arguments: 

- A plane wave wave-function ($Ae^{\pm ipx/\hbar}$) for a free non-relativistic particle is an ideal case that is impossible to conceive in reality unless we make some approximations only for practical purposes. In the case of a plane wave, it is possible to find a reference frame that cancels its momentum ($p=0$), which leads to
an infinite uncertainty on its position, i.e.,$\Delta x=\infty$. However, the existence of a minimum speed $V$ prevents such ideal condition of a plane wave, since $V$ emerges in order to forbid this ideal case of a plane wave ($\Delta p=0$), where the uncertainty on position diverges. In other words, we realize that a minimum speed works like a cut-off for lower speeds by avoiding the existence of rest\cite{N2012}, which leads to a realistic condition where there is no perfect plane wave in reality. Thus we postulate the existence of a non-null minimum speed, so that the momentum of the particle goes to zero when the speed $v$ approaches $V$, but it never reaches $V$. 

Furthermore, here it is important to point out that a
plane wave of a particle with a well-defined momentum and an infinite uncertainty on its position is in fact an idealization, since it is not in agreement with the cosmological reality of a finite universe, whose Hubble radius $R_H\sim 10^{26}$m leads us to think of a maximum uncertainty on position of a particle, but having a finite order of magnitude due to the own finite radius of the universe, as if the particle were free inside a big box ($\Delta x\sim 10^{26}$m), thus having a quasi-zero minimum uncertainty on momentum ($\Delta p_{min}>0$). This implies the presence of a minimum speed in  the scenario of a modified relativity\cite{N2012}. Therefore we can have a quasi-plane wave, but never a perfect plane wave with $\Delta x=\infty$, since the radius of the universe is finite. Thus, as the universe is finite, we can think about a simple model of a particle inside a box (universe) with the order of magnitude of $10^{26}$m instead of the ideal case of a plane wave for a free particle ($\Delta x=\infty$) with a null zero-point energy, which is prevented by the invariance of the minimum speed $V$, being consistent with the realistic case of a particle inside
a finite box by representing a finite universe. 

As the universe has a finite radius, the particle has a zero-point energy, which is in agreement with the 
impossibility of rest. In this sense, we can realize that the absence of rest justified by the minimum speed $V$ further clarifies the understanding of the own uncertainty principle\cite{N2012} in the scenario of a modified space-time due to $V$, which already establishes a (non-null) zero-point energy in the cosmological scenario. 

The quantum-gravitational origin of the fundamental zero-point energy related to a universal minimum speed $V$ indicates that there should be a relationship between $V$ and gravity ($G$), as it was shown in a previous paper, i.e., it was found $V\sim G^{1/2}$\cite{N2016}, $G$ being the constant of gravity.   

The minimum speed $V$ shows us that the luminal particles as the photon ($v=c$) are in equal-footing
with the massive particles ($v<c$) since it is not possible to find a reference frame at rest for any velocity transformations in this modified space-time. Such transformations will be presented later. 

A strong motivation for postulating a minimum speed is the third law of thermodynamics which states that absolute zero temperature is unattainable. From a mechanical point of view, this law prohibits rest. Therefore, in order for mechanics to be compatible with absolute zero, (the third law), an unattainable minimum speed must be postulated by modifying mechanics for very low energies at the quantum level. So, as such a minimum speed is unattainable, it justifies from a mechanical point of view the impossibility of reaching the absolute zero temperature, i.e., the existence of a universal minimum speed can explain the third law 
of thermodynamics. 

In short, let us say that whereas electromagnetism led to a change of paradigm in Newtonian mechanics for very high energies by postulating the speed of light $c$ as an unattainable upper limit of speed (special relativity), thermodynamics leads to a new change of paradigm  in Newtonian mechanics, but now for very low energies, thus leading to the need to 
postulate an unattainable minimum speed $V$.

The dynamics of particles in the presence of a preferred reference frame has a certain similarity with the basic ideas provided by the scenarios of Mach\cite{Mach4}, 
Schr\"{o}dinger\cite{Schroedinger} and Sciama\cite{Sciama}, where such a preferred reference frame
establishes absolutely the inertia and the motions of 
all particles. However, we must stress that the approach of this modified relativity is not machian, as the minimum speed $V$ has a quantum origin associated with the vacuum energy represented by the cosmological constant. 

\section{\label{sec:level1} Modified transformations of space-time and velocity with an invariant minimum speed} 

We review the concepts of reference frames in SSR by introducing the transformations of space-time and velocity in ($1+1$)D space-time. 

Section 4 will show the relationship between the SSR-metric and the dS-metric, having a positive cosmological constant $\Lambda$ (cosmological anti-gravity). This allows us to conclude that SSR provides a background metric that works like a dS-metric. To do that, we will use a spherical model of universe, whose current 
radius is the Hubble radius $R_H$ for the visible 
universe. 

The breaking of Lorentz symmetry at too low energies\cite{N2016} is due to a background field related 
to the ultra-referential $S_V$ (Fig.1). This generates
a modified space-time with a mimimum speed 
$V(=\sqrt{Gm_pm_e}e/\hbar\sim 10^{-14}m/s$)\cite{N2016}, which is the invariant limit of speed for all particles at lower energies. 

The minimum speed $V$ is invariant as is the speed of light $c$. In view of this, $V$ does not change the speed $v$ of any particle for any transformations in 
SSR. 

The ultra-referential $S_V$ is the preferred reference frame in relation to which we have the speeds $v$ of any particles (Fig.1). Therefore, the Lorentz transformations must be changed in the presence of the background reference frame $S_V$ associated with the vacuum energy.  

In the special case below given in $(1+1)D$\cite{N2016}\cite{N2012}, we have found the space-time transformations between the running reference frame $S^{\prime}$ and the preferred reference frame $S_V$ (Fig.1), namely: 

\begin{figure}
\begin{center}
\includegraphics[scale=0.087]{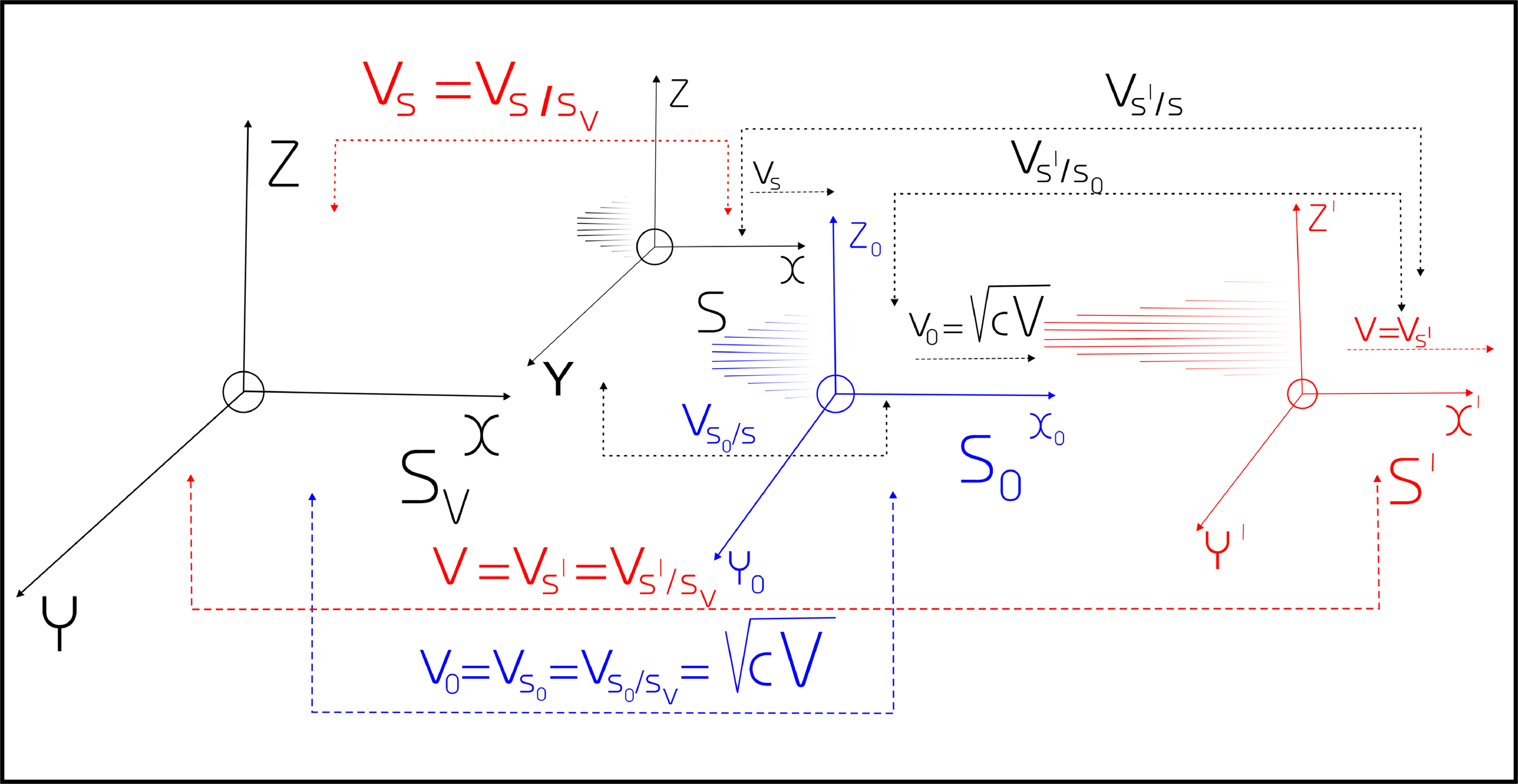} 
\end{center}
\caption{$S^{\prime}$ moves in the axle $x$ with a 
speed $v(>V)$ in relation to the preferred reference frame $S_V$. This figure shows both running 
reference frames $S$ and $S^{\prime}$ with speeds $v=v_S=v_{S/S_V}$ and $v^{\prime}=v_S'=v_{S'/S_V}$,
which are given with respect to the ultra-referential $S_V$. There are two fixed referentials $S_0$ with speed $v_0=v_{S_0/S_V}=\sqrt{cV}$ given with respect to $S_V$ and the ultra-referential $S_V$. So we can obtain the relative velocity between $S'$ and $S$, namely $v_{rel}=v_{S'/S}$, which is presented in Eq.(5) and Eq.(6). This leads to the following important cases, namely:
a) If just the referential $S$ coincides with $S_0$, i.e., $S\equiv S_0$, we get the relative velocity between $S'$ and $S_0$, namely $v_{rel}=v_{S'/S_0}$.
b) If just the referential $S'$ coincides with $S_0$, i.e., $S'\equiv S_0$, we obtain the relative velocity between $S_0$ and $S$, that is to say, $v_{rel}=v_{S_0/S}$.} 
\end{figure}

\begin{equation}
dx^{\prime}=\frac{\sqrt{1-V^2/v^2}}{\sqrt{1-v^2/c^2}}[dX-v(1-\alpha)dt]
\end{equation}

and 

\begin{equation}
 dt^{\prime}=\frac{\sqrt{1-V^2/v^2}}{\sqrt{1-v^2/c^2}}\left[dt-\frac{v(1-\alpha)dX}{c^2}\right], 
 \end{equation}
obtained in refs.\cite{N2016}\cite{N2012}, where
$\alpha=V/v$ and 
$\Psi=\sqrt{1-V^2/v^2}/\sqrt{1-v^2/c^2}$. 

The axes $X$, $Y$ and $Z$ given in the transformations $(1+1)D$ above form the preferred reference frame $S_V$ for the vacuum energy. 

The inverse transformations for $(1+1)D$ were shown in a previous paper\cite{N2016}. If we make the minimum speed $V\rightarrow 0$, we recover Lorentz transformations as a particular case. 

Transformations for the general case $(3+1)D$ were shown in the ref.\cite{N2016}, where it was also demonstrated that the Lorentz and Poincar\'e's groups were broken down. 

The structure of space-time of the Symmetrical Special Relativity (SSR) presents the energy and momentum of a particle, namely $E=m_0c^2\Psi=m_0c^2\sqrt{1-V^2/v^2}/\sqrt{1-v^2/c^2}$\cite{N2016}, in such a way that $E\rightarrow 0$ when $v\rightarrow V$, and $P=m_0v\Psi=m_0v\sqrt{1-V^2/v^2}/\sqrt{1-v^2/c^2}$\cite{N2016}, such that $P\rightarrow 0$ when $v\rightarrow V$. 

For the case $v=v_0=\sqrt{cV}(>V)$, the 
momentum and energy of a particle in SSR is $P_0=m_0v_0=m_0\sqrt{cV}$ and $E_0=E(v_0)=m_0c^2$, as we find $\Psi(v_0)=\Psi(\sqrt{cV})=1$, where the energy $E_0=m_0c^2$ is the same rest energy given in SR. This means that the momentum never vanishes in SSR due to the invariant minimum speed $V$, as there is an intermediary speed $v_0(=\sqrt{cV})$ given with respect to $S_V$ (Fig.1), so that the momentum is non-null ($P_0$) and the energy $E_0(v=v_0)$ is the same rest energy $m_0c^2$ in SR for $v=0$. So, in SSR-theory, $E_0$ is due to the speed $v_0=\sqrt{cV}$ with respect to the ultra-referential $S_V$, since there is no rest in SSR-theory. 

The SSR-metric is a metric with the presence of a conformal factor $\Theta=\Theta(v)=1/(1-V^2/v^2)$\cite{N2016}\cite{Rodrigo1}, thus leading to a Conformal Special Relativity\cite{Rodrigo1} with the presence of an invariant minimum speed $V$, as follows: 

\begin{equation}
dS^{2}=\frac{1}{\left(1-\frac{V^2}{v^2}\right)}[c^2(dt)^2-(dx)^2-(dy)^2-(dz)^2],
\end{equation}
or simply $dS^{2}=\Theta\eta_{\mu\nu}dx^{\mu}dx^{\nu}$, where $\Theta=1/(1-V^2/v^2)$, being $\eta_{\mu\nu}$
the Minkowski metric. 

Dividing Eq.(1) by Eq.(2), we get the velocity transformation, namely: 

\begin{equation}
v_{rel}=v_{S'/S}=\frac{v^{\prime}-v+V}
{1-\frac{v^{\prime}v}{c^2}+\frac{v^{\prime}V}{c^2}}, 
\end{equation}

where $v_{rel}=v_{S'/S}\equiv dx^{\prime}/dt^{\prime}$ and $v^{\prime}\equiv dX/dt$. 

The speed $v^{\prime}=v_{S'}\equiv dX/dt$ is the motion of the reference frame $S^{\prime}$ (Fig.1) with respect to $S_V$ related to the unattainable minimum speed $V$, namely we can write the notation $v_{S'}=v_{S'/S_V}$ for representing the absolute motion of $S^{\prime}$, which is observer-independent, as $S_V$ is the preferred reference frame associated with the vacuum energy. 

The speed $v$ shown in Fig.1 represents the motion of the referential $S$ with respect to the background reference frame $S_V$, i.e., we can write the notation $v=v_{S}=v_{S/S_V}$ as being the absolute motion of $S$ (Fig.1). 

$v_{rel}$ is the relative speed given between the absolute speeds $v_{S'}$ and $v_{S}$, since both of them are given with respect to the preferred frame
$S_V$, namely we have $v_{rel}=v_{S'/S}$ (Fig.1). So we rewrite the transformation in Eq.(4), namely: 

\begin{equation}
v_{S'/S}=\frac{v_{S'/S_V}-v_{S/S_V}+V}{1-\frac{(v_{S'/S_V})(v_{S/S_V})}{c^2}+\frac{(v_{S'/S_V})V}{c^2}}, 
\end{equation}
where $v_{S/S_V}$ is the speed $v$ in $S$ with respect to $S_V$ and $v_{S'/S_V}$ is the speed $v^{\prime}$ 
in $S^{\prime}$ with respect to $S_V$.  

The speed $v_0$ is an intermediary speed, namely $V<<v_0<<c$, so that $\Psi(v_0)=1$. Here we should 
realize that all the speeds $v$, which are not so far from $v_0$ represent the Newtonian approximation in the scenario of SSR, since we get $\Psi(V<<v<<c)\approx 1$. 

Eq.(5) recovers the Lorentz velocity transformation
in the case of $V\rightarrow 0$, where the speeds $v_{S'}$ and $v_S$ would be given with respect to a Galilean reference frame at rest in lab. Thus, in this classical approximation $V\rightarrow 0$, the ultra-referential $S_V$ vanishes and so $v_0$ is zero, namely  $S_0$ would become a Galilean reference frame at rest. 

From Eq.(5), let us consider the following cases, so
that we must consider $v_{S'}\geq v_{S}$, as follows: 

 {\bf a)} If we consider $v_{S'}=c$ (photon) and $v_S\leq c$, we find $v_{rel}=c$, which verifies the invariance of $c$.

 {\bf b)} If we consider $v_{S'}>v_S(=V)$, we find 
 $v_{rel}=``v_{S'}-V"=v_{S'}$. As for instance if we
 have $v_{S'}=2V$ and $v_S=V$, we obtain $v_{rel}=
 ``2V-V"=2V$. This result means that $V$ does 
 not affect the speeds of any particles, since $V$ plays the role of an ``absolute zero of motion''. Thus
 $V$ is invariant, i.e., it has the same value at all directions of the space represented by the isotropic background field, which is related to the preferred reference frame $S_V$. 

 {\bf c)} If we consider $v_{S'}=v_S$, we find
 $v_{S'/S}=``v_S-v_S"=``v-v"=\Delta v=\frac{V}{1-\frac{v^2}{c^2}(1-\frac{V}{v})}$. 

From item ({\bf c}), we consider two special cases, 
namely:

-$c_1$) For $v_S=V$, we find $v_{rel}=``V-V"=V$. In
fact, again we confirm that $V$ is invariant. 

-$c_2$) For $v_S=c$ (photon), we find $v_{rel}=c$. We get the interval $V\leq v_{rel}\leq c$ associated with
the interval $V\leq v_S\leq c$. However, we must have
in mind that there is no massive particle at the preferred reference frame $S_V$, as $V$ 
is unattainable. 

The item $c_2$ shows that there is no rest for the particle on its own reference frame $S$, where 
$v_{rel}$ ($\equiv\Delta v$) is a function that increases with the increasing of $v$. But, if
$V\rightarrow 0$, we always find $v_{rel}\equiv\Delta v=0$, so that it is possible to find rest for $S$. This
is the classical case by recovering the inertial (Galilean) reference frames of SR for $v<c$. 

The inverse transformations from $x^{\prime}\rightarrow X$ and $t^{\prime}\rightarrow t$ for the case $(1+1)D$ and for the general case $(3+1)D$ have already been 
investigated in a previous work\cite{N2016}. 

From the direct transformations above, we can obtain the inverse transformation of velocity, as follows: 

\begin{equation}
v_{S'/S}=\frac{v_{S'/S_V}+v_{S/S_V}-V}{1+\frac{(v_{S'/S_V})(v_{S/S_V})}{c^2}-\frac{(v_{S'/S_V})V}{c^2}},
\end{equation}
where $v_{S'/S_V}=v'$, $v_{S/S_V}=v$ and $v_{S'/S}=v_{rel}$. 

Eq.(6) leads to the relevant cases, namely: 

 {\bf a)} If we consider $v^{\prime}(=v_{S'})=v(=v_S)=V$, we find $``V+V"=V$. Once again we confirm that $V$ is invariant. 
 
 {\bf b)} If we consider $v^{\prime}=v_{S'}=c$ and $v_S\leq c$, we find $v_{rel}=v_{S'/S}=c$, by simply confirming that $c$ is invariant. 
 
 {\bf c)} If we consider $v^{\prime}=v_{S'}>V$ and $v_S=V$, we find $v_{rel}=v_{S'/S}=v_{S'}$. 
 
 From the item {\bf c}, we consider the special cases, as follows: 
 
 -$c_{1}$) If we consider $v^{\prime}=v_{S'}=2V$ and  admitting that $v_S=V$, we find $v_{S'/S}=``2V+V"=2V$. 
 
 -$c_{2}$) If we consider $v^{\prime}=v_{S'}=v_S=v$, we find $v_{S'/S}=``v_S+v_S"=``v+v"=
 \frac{2v-V}{1+\frac{v^2}{c^2}(1-\frac{V}{v})}$.
 
 In the approximation $V<<v<<c$ given for $c_2$, we recover the Newtonian transformation, namely
 $v_{rel}=``v+v"=2v$. 
 
 In the relativistic case, where $v\rightarrow c$, we recover the Lorentz transformation given for such special case $c_2$, i.e., $v^{\prime}=v_{S'}=v_S=v$, where we obtain
 $v_{S'/S}=``v_S+v_S"=``v+v"=2v/(1+v^2/c^2)$. 
 
 \subsection{Do the space-time transformations with an invariant minimum speed form a group?}

 We know that the Lorentz transformations ($L=L(v)$) form a group, as they must obey the conditions, namely:
 
a) $L_2L_1=L(v_2)L(v_1)=L(v_3)=L_3\in L(v)$, which is the closure condition.

b) $L_1(L_2L_3)=(L_1L_2)L_3$, which is the associativity condition.

c) $L_0L=LL_0=L$, such that $L_0=L(0)=I$, which reprersents the identity element. 

d) $L^{-1}L=LL^{-1}=L_0$, being $L^{-1}=L{(-v)}$, which is the inverse element. 

We aim to make an analysis of the new space-time transformations in Eq.(1) and Eq.(2) in order to verify whether they form a group. 

\subsection{Closure condition}

According to Eq.(1) and Eq.(2), we obtain a new matrix of space-time transformation 
($\Lambda$), namely:  

\begin{equation}
\displaystyle\Lambda=
\begin{pmatrix}
\Psi & -\Psi\beta^{*} \\
-\Psi\beta^{*} & \Psi
\end{pmatrix},
\end{equation}
where $\Psi=\frac{\sqrt{1-V^2/v^2}}{\sqrt{1-v^2/c^2}}$. 

Here, we define the notation $\beta^*=\beta\epsilon=\beta(1-\alpha)=(v/c)[1-V/v]$, where 
$\beta=v/c$ and $\alpha=V/v$. 

If we make $V\rightarrow 0$ or $\alpha\rightarrow 0$, we recover the Lorentz matrix, i.e., 
$\Lambda(v)\rightarrow L(v)$, as $\Psi\rightarrow\gamma$ and $\beta^*\rightarrow\beta$. 

Now, we consider $\Lambda_1=\Lambda(v_1)$ as follows: 

\begin{equation}
\displaystyle\Lambda_1=
\begin{pmatrix}
\Psi_1 & -\Psi_1\beta_1^{*} \\
-\Psi_1\beta_1^{*} & \Psi_1
\end{pmatrix}= 
\begin{pmatrix}
\Psi_1 & -\Psi_1\frac{v_1^*}{c} \\
-\Psi_1\frac{v_1^*}{c} & \Psi_1
\end{pmatrix}
\end{equation}

and $\Lambda_2=\Lambda(v_2)$ as being

\begin{equation}
\displaystyle\Lambda_2=
\begin{pmatrix}
\Psi_2 & -\Psi_2\beta_2^* \\
-\Psi_2\beta_2^* & \Psi_2
\end{pmatrix}=
\begin{pmatrix}
\Psi_2 & -\Psi_2\frac{v_2^*}{c} \\
-\Psi_2\frac{v_2^*}{c} & \Psi_2
\end{pmatrix}, 
\end{equation}
so that $\Lambda_2\Lambda_1$ is 

\begin{equation}
\displaystyle\Lambda_2\Lambda_1=[\Psi_2\Psi_1(1+\beta^*_2\beta^*_1)]
\begin{pmatrix}
1 & -\frac{(\beta_1^*+\beta_2^*)}{1+\beta_2^*\beta_1^*} \\
-\frac{(\beta_1^*+\beta_2^*)}{1+\beta_2^*\beta_1^*} & 1
\end{pmatrix},
\end{equation}
where $\beta_1^*=\beta_1\epsilon_1=\beta_1(1-\alpha_1)=(v_1/c)[1-V/v_1]$ and $\beta_2^*=\beta_2\epsilon_2=\beta_2(1-\alpha_2)=
(v_2/c)[1-V/v_2]$.

We obtain that the multiplicative term of the matrix in Eq.(10) is written as 
$\Psi_2\Psi_1(1+\beta^*_2\beta^*_1)=
\sqrt{(1-V^2/v_2^2)(1-V^2/v_1^2)}\frac{1+(v_1^*v_2^*/c^2)}{\sqrt{1-(v_1^2/c^2+v_2^2/c^2-v_1^2v_2^2/c^4)}}$. 

Now we can note that, if the Eq.(10) satisfies the closure condition, Eq.(10) must be equivalent to 

\begin{equation}
\displaystyle\Lambda_2\Lambda_1=\Lambda_3=\Psi_3
\begin{pmatrix}
1  & -\frac{v_3^*}{c} \\
-\frac{v_3^*}{c} & 1
\end{pmatrix}, 
\end{equation}
 where, by comparing Eq.(10) with Eq.(11), we must verify whether the closure condition is satisfied, i.e., $\Psi_3\equiv\sqrt{(1-V^2/v_2^2)(1-V^2/v_1^2)}\frac{1+(v_1^*v_2^*/c^2)}
{\sqrt{1-(v_1^2/c^2+v_2^2/c^2-v_1^2v_2^2/c^4)}}$ and  $v_3^*\equiv(v_2^*+v_1^*)/[1+(v_2^*v_1^*)/c^2]$. However, we first realize that such
 speed transformation, which must be obeyed in order to satisfy the closure condition, differs from the correct speed transformation 
(Eq.(6)) that has origin from the new space-time transformations given in Eq.(4), Eq.(5) and Eq.(6). Thus, according to Fig.1, 
if we simply redefine $v^{\prime}=v_2$ and $v=v_1$, we rewrite the correct transformation (Eq.(6)) as being
$v_{rel}=v_3=(v_2+v_1^*)/[1+(v_2v_1^*)/c^2]$, where $v_1^*=v_1-V$. Now, we note that the correct transformation for $v_3$ (Eq.(6)) is not the
same transformation given in the matrix above (Eq.(10)), i.e., we get $v_3\neq (v_2^*+v_1^*)/[1+(v_2^*v_1^*)/c^2]$. 

One of the conditions of the closure relation is that the components outside the diagonal of the matrix in Eq.(10) must include
$v_3$ given by Eq.(6), which does not occur. Therefore, we conclude that such condition is not obeyed in the spacetime
with a minimum speed associated with a preferred reference frame, i.e., we have 
$\Lambda_2\Lambda_1\neq\Lambda_3$, which does not form a group. In order to clarify further this question, let us make the approximation $V=0$ or $v_1\gg V$ and $v_2\gg V$ in
Eq.(10), so that we recover the closure relation of the Lorentz group as a special case, namely: 

\begin{equation}
\displaystyle(\Lambda_2\Lambda_1)_{V=0}=L_2L_1=
\displaystyle L_3=\gamma_3
\begin{pmatrix}
 1  &     -\frac{v_3}{c}\\
 -\frac{v_3}{c}  & 1
\end{pmatrix}, 
\end{equation}
which is the closure condition of the Lorentz group, where 
$v_3=(v_1+v_2)/[(1+(v_1v_2)/c^2]$ and $\gamma_3=1/\sqrt{1-v_3^2/c^2}$. 

Now it is clear that the Lorentz transformation of speeds is outside 
the diagonal of the matrix in Eq.(12), i.e., we have $v_3=(v_1+v_2)/[1+(v_1v_2)/c^2]$.
In order to finish the verification of the
closure condition above, we just verify that the multiplicative term of the matrix (Eq.(12)) is $\gamma_2\gamma_1(1+\beta_2\beta_1)=
\left(1+\frac{v_1v_2}{c^2}\right)/\sqrt{1-\left(\frac{v_1^2}{c^2}+\frac{v_2^2}{c^2}-\frac{v_1^2v_2^2}{c^4}\right)}=\gamma_3$. To do this,
we consider $v_3=(v_1+v_2)/[1+(v_1v_2)/c^2]$, so that we use this transformation to be inserted into $\gamma_3=1/\sqrt{1-v_3^2/c^2}$. We show that $\gamma_3=1/\sqrt{1-v_3^2/c^2}=\gamma_2\gamma_1(1+\beta_2\beta_1)$. However, by starting from this same procedure
for obtaining $\Psi_3=\frac{\sqrt{1-V^2/v_3^2}}{\sqrt{1-v_3^2/c^2}}$, where we need to use the correct transformation for $v_3$ (Eq.(6)),
we find $\Psi_3\neq\Psi_2\Psi_1(1+\beta_2^*\beta_1^*)$ and thus we definitively conclude that the closure condition does not apply to the spacetime transformations of SSR. 

\subsection{Identity element} 

In the case $(1+1)D$ spacetime, the Lorentz group generates the identity element $L_0=I_{(2X2)}$, since $L_0L=LL_0=L$. As the Lorentz matrix 
is $\displaystyle L_{(2X2)}=\begin{pmatrix}
 \gamma  &     -\beta\gamma \\
-\beta\gamma  & \gamma
\end{pmatrix}$, for $v=0$ or $\beta=0$ (rest condition), we recover the identity matrix $\displaystyle
 I_{(2X2)}=\begin{pmatrix}
 1 &     0 \\
 0  &    1
\end{pmatrix}$, since $\gamma_0=\gamma(v=0)=1$. This trivial condition of rest and also
the fact that $det(L)=1$ (rotation matrix) guarantee the equivalence of rest and inertial motion. Acceleration would be capable of breaking this equivalence, as well as the 
spacetime with an invariant minimum speed, where rest is prohibited.

The new transformations are represented by the matrix $\displaystyle\Lambda=
\begin{pmatrix}
\Psi & -\beta(1-\alpha)\Psi \\
-\beta(1-\alpha)\Psi & \Psi
\end{pmatrix}$, where we have $\beta^*=\beta\epsilon=\beta(1-\alpha)$, with $\alpha=V/v$. Now, we should notice that there is no speed $v$
that generates the identity matrix from the new matrix. We expect that the hypothesis $v=V$ could do that, however,
if we make $v=V$ ($\alpha=1$), we obtain the null matrix, i.e., $\displaystyle\Lambda(V)=\begin{pmatrix}
0 & 0\\
0 & 0
\end{pmatrix}$, since $\Psi(V)=0$. So, we get $\Lambda(V)\Lambda=\Lambda_V\Lambda=\Lambda\Lambda_V=\displaystyle\begin{pmatrix}
0 & 0\\
0 & 0
\end{pmatrix}\neq\Lambda$, where we have $\Lambda=\Lambda(v>V)$. 

We conclude that there is no identity element in the spacetime of SSR. This means that
there should be a distinction of motion and rest, as there is a preferred reference frame
associated with an invariant minimum speed that breaks down the Galilean notion of the 
equivalence between inertial motion and rest. In this sense, the own minimum speed in subatomic level prevents rest. Hence the existence of a zero-point energy associated with 
the vacuum energy, which leads to the cosmological constant within a cosmological scenario. 

\subsection{Inverse element}

We know that the inverse element exists in Lorentz transformations that form a group, i.e., we have $L^{-1}(v)=L(-v)$. This means
that we can exchange the observer in the reference frame $S$ at rest by another observer in the reference frame $S^{\prime}$ with speed $v$ in 
respect to $S$, so that the other observer at $S^{\prime}$ observes $S$ with a speed $-v$. This symmetry has origin from the galilean
relativity, which it is due to the equivalence of rest and inertial motion. Here we must stress that such equivalence is broken down in the spacetime transformations of SSR, as the  minimum speed related to a preferred reference
frame $S_V$ (Fig.1) introduces a preferential motion $v(>V)$ that cannot be exchanged by 
$-v$, since rest does not exist in the spacetime of SSR. In view of this, we show that 
$\Lambda^{-1}(v)\neq\Lambda(-v)$, such that we obtain $\Lambda(-v)\Lambda(v)=\displaystyle\theta^2
\begin{pmatrix}
\gamma & \beta(1-\alpha)\gamma \\
\beta(1-\alpha)\gamma & \gamma
\end{pmatrix}$ $\times$
$\displaystyle\begin{pmatrix}
\gamma & -\beta(1-\alpha)\gamma \\
-\beta(1-\alpha)\gamma & \gamma
\end{pmatrix}=\displaystyle\Psi^2\begin{pmatrix}
 \left(1-\frac{v^{*2}}{c^2}\right) & 0 \\
 0 & \left(1-\frac{v^{*2}}{c^2}\right)
\end{pmatrix}\neq I_{(2X2)}$. For $V=0$ ($\alpha=0$), we recover the inverse element 
$I_{2X2}$ of the Lorentz group, which is a rotation group.

Therefore, we have shown that the transformations of SSR do not form a group. We have also provided a physical explanation for such Lorentz
violation due to an invariant minimum speed that breaks down the equivalence of rest and inertial motion. 

We have concluded that the matrix $\Lambda$ does not represent a rotation matrix. Thus,
we realize that the new transformations are not represented by the rotation group $SO(3)$ (Lie group), whose elements
$R(\vec\alpha)$ and $R(\vec\beta)$ must obey the closure condition $R(\vec\alpha)R(\vec\beta)=R(\vec\gamma)$, such that $\vec\gamma=
\gamma(\vec\alpha,\vec\beta)$, $R(\vec\gamma)$ being a new rotation that belongs to the group. So we find $det(R)=+1$ (rotation condition), whereas we obtain $det(\Lambda)=\theta^2\gamma^2\left[1-\frac{v^2(1-\alpha)^2}{c^2}\right]$, where $0<det(\Lambda)<1$, violating the rotation condition. 

Finally, we realize that there could be a more complex mathematical structure in order to encompass the new transformations. Such a mathematical structure that does not form the group $S0(3)$ should be deeply investigated elsewhere.

 \section{\label{sec:level1} The correspondence of the SSR-metric with the dS-metric by means of the 
 cosmological constant} 

\begin{figure}
\begin{center}
\includegraphics[scale=0.27]{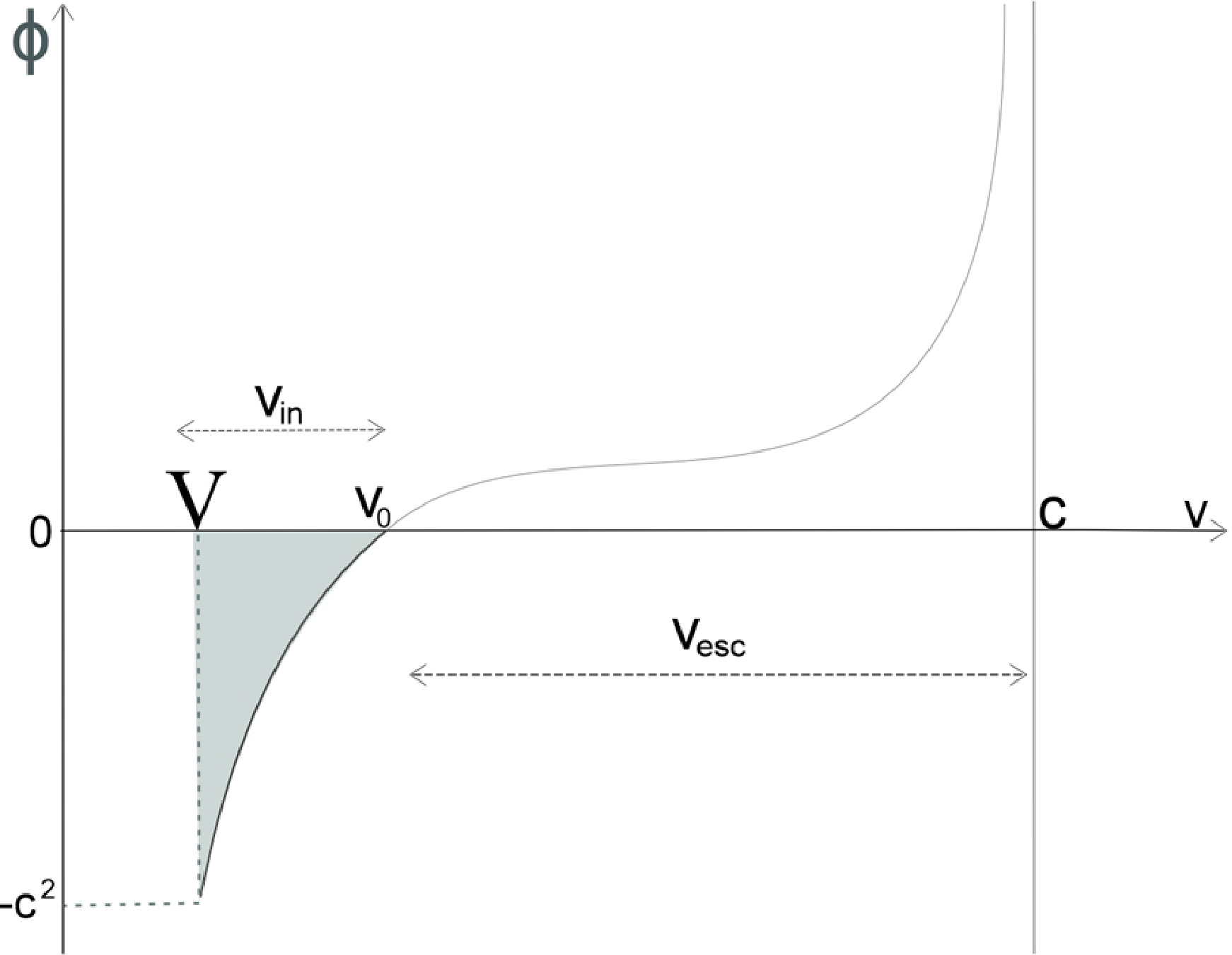}
\end{center}
\caption{The figure represents the potential
$\phi(v)=c^{2}\left(\sqrt{\frac{1-\frac{V^2}{v^2}}{1-\frac{v^2}{c^2}}}-1\right)$ shown in Eq.(14). It presents two phases given by gravity at the right side and anti-gravity at the left site. The barrier at the right side is the relativistic limit given by speed of light $c$,  where $\phi\rightarrow\infty$. On the other hand, the barrier at the left side is the anti-gravitational limit given by SSR. The minimum speed $V$ is related to the potential $\phi(V)=-c^2$. There is an intermediary interval of speeds so-called the Newtonian regime ($V<<v<<c$), where, for $v=v_0=\sqrt{cV}$, occurs a phase transition from gravity to anti-gravity.}
\end{figure}

Let us show that the SSR-metric [Eq.(3)] corresponds to a dS-metric (positive cosmological constant), since the universal minimum speed $V$ related to the ultra-referential $S_V$ shoud be associated with a positive cosmological constant $\Lambda$. Therefore, a conformal factor\cite{Rodrigo1} emerges from 
the SSR-metric, so that it depends on $\Lambda$\cite{Rodrigo1}. In order to do that, let us build a model of a spherical universe with a Hubble radius $R_H$ and a vacuum energy density $\rho_{\Lambda}$. 
 
The surface of the sphere, which represents the frontier of the visible universe with the objects like galaxies, etc experience an anti-gravity given by the accelerated expansion of the universe. This anti-gravity is due to the vacuum energy related to a dark
mass inside the Hubble sphere. 

Each galaxy on the surface of the sphere works like a proof body that interacts with such sphere with a dark mass $M_{\Lambda}$. This interaction is a simple case of interaction between two bodies. Thus, let us show that there is an anti-gravitational interaction between the proof mass $m_0$ on the surface of the dark sphere and the own dark sphere with a dark mass $M_{\Lambda}=M$. 

To investigate the anti-gravity between the proof mass $m_0$ and the dark mass $M$ of the Hubble sphere, we first consider the model of a proof particle with mass $m_0$ that escapes from a gravitational potential
$\phi$ on the surface of a sphere of matter with mass $M_{matter}$, i.e., we write $E=m_0c^2(1-v^2/c^2)^{-1/2}\equiv m_0c^2(1+\phi/c^2)$.

$E$ is the escape relativistic energy of the proof particle with mass $m_0$ and $\phi=GM_{matter}/R$, $R$ being the radius of the sphere of matter. In such classical case, the interval of escape velocity ($0\leq v<c$) is related to the interval of potential ($0\leq\phi<\infty$), where we consider $\phi>0$ to be the attractive gravitational potential. 

The breaking of Lorentz symmetry in SSR is due to the presence of the preferred reference frame $S_V$ related to the vacuum energy inside the dark sphere. This energy has origin from a non-classical gravity, 
thus leading to a repulsive gravitational potential,
which is negative as it represents anti-gravity, i.e., $\phi<0$ (Fig.2). 

According to SSR, in the model of spherical universe, we write the deformed relativistic energy of the
proof particle with mass $m_0$, namely: 

\begin{equation}
E=m_0c^2\left(1+\frac{\phi}{c^2}\right)=
m_0c^2\left(\frac{{1-\frac{V^2}{v^2}}}{{1-\frac{v^2}{c^2}}}\right)^{\frac{1}{2}},
\end{equation}
from where we get
\begin{equation}
\phi=\phi(v)=\left[\left(\frac{{1-\frac{V^2}{v^2}}}{{1-\frac{v^2}{c^2}}}\right)^{\frac{1}{2}}-1\right]c^2. 
\end{equation} 

We realize that Eq.(14) shows us two situations, as follows: 

i) The Lorentz sector is $\phi=(\gamma-1)c^2$. It is the sector of gravity, as the sphere $M$ is made of attractive matter. In this classical case, the speed $v$ represents the escape velocity ($v_{esc}$). 

ii) The anti-gravity sector is $\phi=\phi_q=(\theta-1)c^2$. The factor $\theta=(1-V^2/v^2)^{1/2}$ is represented by a dark sphere with mass $M$. The speed $v$ should be interpreted as the input speed ($v_{in}$), namely the speed of a proof particle that escapes from anti-gravity. 

SSR prevents rest of a particle based on Eq.(14). So, we must notice that $v$ cannot be zero even in the absence of potential $\phi$ ($\phi=0$), i.e, we find $v=v_0=\sqrt{cV}$, so that $\phi(v_0)=0$ 
in Eq.(14) (Fig.2). 

The absence of gravitational potential ($\phi=0$) at the point $v=v_0$ leads us to conclude that $v_0$ 
is the only velocity that means both of the escape and input velocities of a particle, since $v_0$ is a point of transition between gravity and anti-gravity. 

In sum, we have two sectors of gravitational potential, namely the classical (matter) and quantum (vacuum) sectors, as follows: 

\begin{equation}
\phi=\left\{
\begin{array}{ll}
\phi_{q}:&\mbox{$-c^2<\phi\leq 0$ for $V< v\leq v_0$}.\\\\
\phi_{m}:&\mbox{$0\leq\phi<\infty$ for $v_0\leq v<c$}, 
\end{array}
\right.
\end{equation}
where $v_0$ for $\phi=0$ is the point of phase transition between gravity with $\phi>0$ for $v>v_0$ and anti-gravity governed by vacuum with  
$\phi<0$ for $v<v_0$. 

The transition point $v_0$ given in relation to the 
ultra-referential $S_V$ is also an observer-independent velocity. Thus there is no observer at 
the ultra-referential $S_V$.   

According to Fig.2, we see that $\phi=-c^2$ is the most fundamental vacuum energy associated with $S_V$,  when making $v=v_{in}=V$ in Eq.(14). So we get
$\phi(V)=-c^2$.  

By considering the spherical universe with a Hubble radius $R_H(=R_u)$ and a vacuum energy density 
$\rho_{\Lambda}$, we find the vacuum energy inside the sphere, namely $E_{dark}=\rho_{\Lambda} V_u=Mc^2$. $V_u$ is the spherical volume of the observable universe and $M$ is the total dark mass for the vacuum energy. 

We already know that $\rho_{\Lambda}$ is too low. In 
view of this, we conclude that the Newtonian gravitational potential (a weak potential) should be a very good approximation for representing the spherical model for the visible universe. Thus, we obtain the repulsive gravitational potential 
$\phi_q<0)$ on the surface of the Hubble sphere, as follows: 

\begin{equation}
\phi_q=-\frac{GM}{R_u}=
-\frac{4\pi G\rho R_u^2}{3c^2}=-\frac{G\rho V_u}{R_uc^2},
\end{equation}
where $M=\rho V_u/c^2$, $\rho=\rho_{\Lambda}$ is the vacuum energy density and $V_u(=4\pi R_u^3/3)$ is the Hubble volume.

We have $\rho_{\Lambda}=\Lambda c^2/8\pi G$. Thus by substituting $\rho_{\Lambda}$ 
in Eq.(16), we get
the repulsive potential, as follows: 

\begin{equation}
\phi_q=-\frac{\Lambda R_u^2}{6},
\end{equation}
where $R_u\sim 10^{26}$m. 

The universe is practically governed by the vacuum energy ($73\%$ of its total mass). Now consider a 
proof mass over the spherical surface $4\pi R_H^2$, so that it experiences a cosmological anti-gravity. In order to overcome such an anti-gravity by escaping into the sphere, the proof particle needs to have 
an input spped $v_{in}$.  

Eq.(14) provides the input speed $v_{in}$ in the 
anti-gravitational regime for $v<v_0$, namely the factor $(1-V^2/v^2)^{1/2}$ in Eq.(14) prevails in the 
case of a repulsive potential $\phi$ (anti-gravity).
So, in this non-classical case ($\phi<0$), we can neglect the Lorentz factor $\gamma$, i.e., the attractive sector, where $\phi>0$. 

By neglecting $\gamma=(1-v^2/c^2)^{-1/2}$ in Eq.(14), we can compare the anti-gravity sector with Eq.(17) given for a radius $r(=ct)$, namely $\phi_q(=-\Lambda r^2/6)$, so that we get $\phi_q/c^2$, as follows: 

\begin{equation}
\frac{\phi_q}{c^2}=-\frac{\Lambda r^2}{6c^2}=\left(1-\frac{V^2}{v^2}\right)^{\frac{1}{2}}-1, 
\end{equation}
where $\phi_q$ is the repulsive potential.

From Eq.(18), we can get the scale factor $\Theta(v)$ in its equivalent forms, namely:

\begin{equation}
\Theta(v)=\frac{1}{\left(1-\frac{V^2}{v^2}\right)}\equiv\frac{1}{\left(1+\frac{\phi_q}{c^2}\right)^2}\equiv\frac{1}{\left(1-\frac{\Lambda r^2}{6c^2}\right)^2}, 
\end{equation} 

By replacing $\Theta(v)$ in Eq.(3) by its equivalent form that depends on $\Lambda$, i.e., 
$\Theta(\Lambda)$, we can write SSR-metric, as follows: 

\begin{equation}
dS^{2}=\frac{1}{\left(1-\frac{{\Lambda}r^2}{6c^2}\right)^2}[c^2(dt)^2-(dx)^2-(dy)^2-(dz)^2], 
\end{equation} 

or simply 

\begin{equation}
dS^{2}=\Theta(\Lambda)\eta_{\mu\nu}dx^{\mu}dx^{\nu}, 
\end{equation}
where $\eta_{\mu\nu}$ is the Minkowski metric and
$\mathcal G_{\mu\nu}=\Theta(\Lambda)\eta_{\mu\nu}$. 

In Eq.(20), by making $\Lambda=0$, we find 
$\Theta=1$, so that the Minkowski metric 
$\eta_{\mu\nu}$ is recovered. 

We have $\Lambda=-6\phi_q/r^2$ [Eq.(17)]. So, for
$r\rightarrow\infty$ ($\Lambda\rightarrow 0$), the interval $d S^{2}$ recovers the Minkowski interval $ds^2$, namely $dS^{2}\rightarrow ds^2=\eta_{\mu\nu}dx^{\mu}dx^{\nu}$\cite{jgpereira}. 

We realize that Eq.(20) is a dS-metric with 
$\Lambda>0$, as we must have $\phi<0$. 

Eq.(17) and Eq.(14) leads us to conclude that a cosmological constant $\Lambda$ emerges from SSR, namely $\Lambda=-6\phi_q/r^2$ [Eq.(17)].

Now it is easy to realize that SSR has a deep correspondence with the de-Sitter (dS) relativity\cite{jgpereira} shown by Eq.(20). 

In order to get the tiny order of magnitude of 
$\Lambda$, we first consider $\Lambda$ [Eq.(17)] given for the Hubble radius $R_H(\sim 10^{26}m)$, so that we obtain 

\begin{equation}
\Lambda=\Lambda(R_H)=-\frac{6\phi_q}{R_H^2},
\end{equation}
where $r=R_H$. 

The repulsive potential $\phi_q$ is within the following range, namely $-c^2\leq\phi_q\leq 0$ shown
in Fig.2. 

Finally, if we admit the most repulsive potential 
$\phi_q=-c^2$ associated with $V$ at the ultra-referential $S_V$, we get  

\begin{equation}
\Lambda=\frac{6c^2}{R_H^2}\sim 10^{-35}s^{-2}. 
\end{equation}  

\section{\label{sec:level1} The cosmological constant in the zero-gravity limit ($z=1$) according to the Boomerang experiment} 

The very small $\Lambda(\sim 10^{-35}s^{-2})$ may have implication in a realistic cosmological scenario of a flat space-time governed by a dark energy 
($\Omega_{\Lambda}\approx 0.73$) according to the Boomerang experiment\cite{boomerang}\cite{boomerang1}, which 
is consistent with the $\Lambda CDM$ scenario. In order to realize such an implication, we first need to approximate the metric given in Eq.(20) to a quasi-flat metric representing a universe with a slightly accelerated expansion, so that we should make $\Lambda\approx 0$ or even $\Lambda<<1$ in Eq.(20). In doing that, we get the approximation for a very weak potential , i.e., we make the approximation $\Lambda r^2/6<<c^2$ in Eq.(20). Thus, we write $\Theta=(1-\Lambda r^2/6c^2)^{-2}\approx(1-\Lambda r^2/3c^2)^{-1}$, so that we obtain the following metric: 

\begin{equation}
d\mathcal S^{2}=\frac{1}{\left(1-\frac{{\Lambda}r^2}{3c^2}\right)}[c^2(dt)^2-(dx)^2-(dy)^2-(dz)^2], 
\end{equation}
from where we can get the cosmological constant within the scenario of a very weak anti-gravity. In order to do that, we consider the most fundamental vacuum at $S_V$, so that we make $\phi=-c^2$ in Eq.(22). Thus we obtain $\Lambda$ in function of the Hubble time, namely: 

\begin{equation}
\Lambda=3c^2/r^2=3/\tau^2, 
\end{equation}
with $r=c\tau$, $\tau$ being a certain Hubble time. 

We know that $\Lambda=8\pi G\rho/c^2=k\rho$, where $8\pi G/c^2$ is 
the well-known constant $k$ in the Einstein equation. So we can write
$k\rho=3/\tau^2$, where we find $\rho=3/k\tau^2=\Lambda/k$.

Now, we stress that there must be a critical density $\rho_c=3/k\tau_0^2$ given exactly in the zero-gravity limit associated with the Hubble time 
$\tau_0$ at which the universe went over from a decelerating to an accelerating expansion, thus  obtaining the numerical value $\tau_0$, which allows us to get the numerical value of $\Lambda_0(=k\rho_c=3/\tau_0^2)$ in agreement with measurements\cite{boomerang}\cite{boomerang1}. 

In a previous work\cite{Lambda}, it was shown that 
the current universe is in an accelerated expansion. In spite of such theory\cite{Lambda} does not 
present a cosmological constant, it predicts 
an accelerated expansion of the universe. Thus, it is equivalent to the general relativity (GR) with 
a positive cosmological constant. 

When we calculate the zero-zero component of the field equations ($R_{\mu\nu}-(R/2)g_{\mu\nu}=kT_{\mu\nu}$) in the framework of this theory\cite{Lambda}, we find   

\begin{equation}
R^0_0-\frac{1}{2}\delta^0_0 R=k\rho_{eff}=k(\rho-\rho_c),
\end{equation}
with $\rho_c=3/k\tau_0^2=\Lambda_0/k$\cite{Lambda} being the critical density in the zero-gravity 
limit and $\tau_0$ is Hubble's time in this 
same limit. 

From Eq.(26), we notice that the effective density 
$\rho_{eff}=0$ only if $\rho=\rho_c$, which corresponds exactly to the zero-gravity limit. 

Here we must stress that the framework of this theory\cite{Lambda} given 
by Eq.(26) is consistent with Eq.(24), since both equations provide information about the existence of a Hubble time 
$\tau_0$ at the zero-gravity limit. Such consistency is not surprising, since in a previous 
paper\cite{Rodrigo1} it has already been proven that the conformal metric of SSR is a solution of the Einstein equation with $\Lambda>0$. 

The framework that leads to Eq.(26)\cite{Lambda} uses a Riemannian four-dimensional formulation of gravitation. In this formulation, the Hubble coordinates are used, namely distances and velocity instead of space and time. 

When solving the field equations, there emerge
three possibilities, namely a decelerating
expansion of the universe ($\Omega_m>1$), followed by a constant expansion ($\Omega_m=1$) and after an accelerating expansion ($\Omega_m<1$). 

Since we are interested only in the latter phase of acceleration described by the background dS-metric in Eq.(24), then for the accelerating phase given by the theory\cite{Lambda}, we find  

\begin{equation}
 H_0=h[1-(1-\Omega_m)z^2/6], 
\end{equation}
where $\Omega_m=\rho_m/\rho_c(<1)$ and 
$h=\tau^{-1}$.  

The parameter $z$ is the redshift, which determines the distance. In turn, this distance is associated 
with $H_0$. So, at the zero-gravity limit, by choosing $z=1$ and having $\Omega_m= 0.245$ 
its current value given in the Table 1 in ref.\cite{Lambda}, then from Eq.(27) we get
$H_0=0.874h$. 

For $z=1$, we have the Hubble parameter $H_0=70km/s-Mpc$, and thus $h=h_0=80.092km/s-Mpc$. So, we get 
$\tau=\tau_0=3.938\times 10^{17}s$, which is exactly the numerical value of the Hubble time 
$\tau_0$ at the zero-gravity limit. 

 By substituting the Hubble time $\tau_0$ above in Eq.(25), we obtain 
 
 \begin{equation}
 \Lambda_0=\frac{3}{\tau_0^2}=1.934\times 10^{-35}s^{-2}.
 \end{equation}
 
 In sum, we realize that the SSR-theory has proposed a modified space-time with an invariant minimum speed $V$ associated with the preferred reference frame $S_V$. Therefore, SSR amplifies the framework of the special relativity (SR), which leads to provide the first principles for understanding the tiny value of the cosmological constant, being in  agreement with the observational data\cite{boomerang}\cite{boomerang1}
 \cite{15,16,17,18,19,20,21} by avoiding the renormalization procedures of the quantum vacuum in the Quantum Field Theories (QFT), which lead to the well-known {\it Cosmological Constant Problem}\cite{Lambda1}\cite{Lambda2}\cite{Lambda3}. So now we understand better that the present theory advances the current
space-time paradigm of relativity, by making it possible to solve the cosmological constant puzzle.
 
\section{The Weyl Geometrical Structure of SSR}
 
An important aspect of Weyl geometric structure is that when one goes from one frame $(M, g, \sigma)$ to another frame 
$(M, \bar{g}, \bar{\sigma})$\cite{Fonseca} by using the gauge transformations \cite{Rosen}\cite{Romero}, we have
 
\begin{equation} \label{Weyl}
\begin{aligned}
\bar{g} &= e^{f}g \\
\bar{\sigma} &= \sigma + df, 
\end{aligned}
\end{equation}
the affine geodesic curves, where $f$ is a scalar function\cite{Sanomiya} defined in the differentiable manifold $M$, with a metric $g$ and the Weyl field $\sigma$ one keeps unaltered\cite{Edgar}. In a certain sense we can characterize the Weyl geometry as an extension of Riemannian geometry where the covariant derivative of the Metric Tensor $g$ is given by\cite{Madriz} 

\begin{equation}\label{affine}
\nabla_\alpha g_{\beta \lambda} = \sigma_{\alpha}g_{\beta \lambda},
\end{equation}
with $\sigma_\alpha$ being the components of a one-form field $\sigma$ in a local coordinate basis. In this case we can write the affine connection as\cite{Romerob}

\begin{equation}
\Gamma^{\alpha}_{\beta \lambda} = \left\{ \begin{array}{c} k \\ ij \end{array} \right\} -\frac{1}{2}g^{\alpha \mu}[g_{\mu\beta} \sigma_{\lambda} + g_{\mu\lambda} \sigma_{\beta}-g_{\beta \lambda} \sigma_\mu],
\end{equation}
where $\left\{ \begin{array}{c} k \\ ij \end{array} \right\}$ is the usual Christoffel symbols.\\

Therefore, the equations of motion, which are the geodesic equations of this geometric 
structure\cite{Ibraginov}

\begin{equation}\label{geo}
\frac{d^2 x^{\alpha}}{d\tau^2} + \Gamma^{\alpha}_{\beta \lambda}\frac{dx^{\beta}}{d\tau}\frac{dx^{\lambda}}{d\tau} = 0
\end{equation}
are invariant, with $\tau$ being the usual proper time.

The beauty of Weyl's treatment is that under the condition in Eq.(\ref{affine}) the connection $\nabla_{\alpha}$ and therefore the geodesic equations are invariant under transformations in Eq.(\ref{Weyl}). On the other hand consider two vector fields $V$ and $U$ which can be parallel-transported along a curve $\alpha = \alpha(\lambda)$. So from 
Eq.(\ref{affine}) one can written\cite{Lobo} 

\begin{equation}\label{aaffine}
\frac{d}{d\lambda}g(V, U)=\sigma(\frac{d}{d\lambda})g(V,U), 
\end{equation}
with $\frac{d}{d\lambda}$ being the vector tangent to $\alpha$. Integrating 
Eq.(\ref{aaffine}) along the curve $\alpha$ from a point $P_0 = \alpha(\lambda_0)$ one obtains

\begin{equation}\label{iaaffine}
g(V(\lambda), U(\lambda)) = g(V(\lambda_0), U(\lambda_0))e^{\int^{\lambda}_{\lambda_0}\sigma(\frac{d}{d\rho})\rho}. 
\end{equation}

For a specific case where $U = V$ and $L(\lambda)$ is the length of the vector $V(\lambda)$ at a point $P = \alpha(\lambda)$ of the curve, so in the coordinate system ${x^{\alpha}}$, the Eq.(\ref{aaffine}) becomes 

\begin{equation}\label{length}
\frac{dL}{d\lambda} = \frac{\sigma_\alpha}{2}\frac{dx^{\alpha}}{d\lambda}L.
\end{equation}

Considering a closed curve with the conditions $\alpha \ : [a,b] \ \in \ R \rightarrow M $\cite{Dahia} or in a concise way $\alpha(a) = \alpha(b)$ the both concepts in 
Eq.(\ref{iaaffine}) and Eq.(\ref{length}) conduct to

\begin{equation}\label{flength}
L = L_0e^{\frac{1}{2}\oint \sigma_\alpha dx^{\alpha}}, 
\end{equation}
with $L_0$ and $L$ being respectively the values of $L(\lambda)$ at $a$ and $b$.

The theoretical consequence of Eq.(\ref{flength}) is that applying the Stokes theorem we obtain the expression 

\begin{equation}\label{flestrength}
L = L_0e^{-\frac{1}{4}\int \int F_{\mu \nu}dx^{\mu} \wedge dx^{\nu}}, 
\end{equation}
so that now we can define $F_{\mu\nu} = \partial_{\nu}\sigma_{\mu}-\partial_{\mu}\sigma_{\nu}$. The result that one obtains is that a vector have its original length preserved if the 2-form $F = d\sigma=\frac{1}{2} F_{\mu\nu} dx^{\nu} \wedge dx^{\mu}$ vanishes.
\
\subsection{\label{sec:level1} Weyl Conformally Flat Spacetimes and Gravity}

We wish to show in a simple and direct way that the factor $\Theta(v)$ in Eq.(3) is a conformal Weyl factor and that the SSR-theory has a Weyl conformal geometry by verifying its Newtonian weak-field limit in the scenario of SSR, i.e., $V<<v<<c$, namely we have the approximation\cite{CNassif}

\begin{equation}
g_{\mu\nu}\approx\eta_{\mu\nu} + \epsilon h_{\mu\nu}, 
\end{equation}
where $\eta_{\mu\nu}$ is the Minkowski tensor, $\epsilon$ is a small parameter and $\epsilon h_{\mu\nu}$ is a small time-independent perturbation on the metric tensor 
$g$\cite{LPuncheu}. 

Now by considering a conformally flat spacetime we can write

\begin{equation}
g_{\mu\nu} = e^{\phi}\eta_{\mu\nu}\backsimeq (1 + \phi/c^2)\eta_{\mu\nu},
\end{equation}
where it is interesting to observe that for $\Lambda<<1$, which means 
$\phi/c^2<<1$ or $V<<v<<c$ (Newtonian limit in SSR) in Eq.(3), this leads to the aproximation $g_{\mu\nu}\approx (1+\phi/c^2)\eta_{\mu\nu}\approx{(1+\Lambda r^2/3c^2)}\eta_{\mu\nu}$ ($\Lambda<<1$), which is valid for redshift $z\approx 1$ with a very weak antigravity (almost flat space-time) with a slight cosmic acceleration, from where we have obtained the very small value of $\Lambda=\Lambda_0=1.934\times 10^{-35}s^{-2}$ in the approximation of Euclidian space or zero-gravity (quasi-flat space). 

Therefore, in such regime of transition from gravity to anti-gravity\cite{NassifPramana}, the Minkowski metric becomes slightly deformed, namely:

\begin{equation}\label{Min}
ds^2\approx (1 + \phi/c^2)[(dx^0)^2-(dx^1)^2]-(dx^2)^2-(dx^3)^2], 
\end{equation}
with $x^0 = ct$ and $\phi/c^2<<1$ (almost flat space-time). 

If we consider the motion of a test particle in the spacetime (Eq.(\ref{Min})), as 
$\Gamma^{\mu}_{\alpha \beta}$ is invariant under Eq.(\ref{Weyl}), the following approximation can be realized

\begin{equation}\label{approximation}
\left(\frac{ds}{dt}\right)^2 \backsimeq c^2(1 + \epsilon h_{00}) = 
c^2(1 + \phi/c^2).
\end{equation}

So the geodesic equation is

\begin{equation}
\frac{d^2x^{\mu}}{ds^2} + \Gamma^{\mu}_{\alpha\beta}\frac{dx^{\alpha}}{ds}\frac{dx^{\beta}}{ds}, 
\end{equation}
which reduces itself to 

\begin{equation}
\frac{d^2x^{\mu}}{dt^2} + \Gamma^{\mu}_{00}\left(\frac{dx^0}{ds}\right)^2, 
\end{equation}
which in the approximation in Eq.(\ref{approximation}) may be written as 

\begin{equation}
\frac{d^2x^{\mu}}{dt^2} + c^2\Gamma^{\mu}_{00}
\end{equation}

Thus we can observe that the conformal factor in Eq.(3) includes the same extended geometrical structure and the same weak-field Newtonian limit, i.e., we have $V<<v<<c$ or $\phi/c^2<<1$. 

It is worth mentioning that the Weyl structure is overlying to Riemannian geometry and the fact that the field equations are invariant under the frame transformations\cite{Karim} occasioned by the transformations (\ref{Weyl}), in a certain proportion show that the geometric structure of the SSR can address fundamental aspects of nature such as the cosmological constant
(vacuum energy)\cite{NassifPramana} and the dark energy.

\subsection{Differential Geometry}

The central part of our analysis is to consider that

\begin{equation}\label{coint}
\oint\sigma\left(\frac{d}{d\lambda}\right)d\lambda = 0
\end{equation}
and this can be achieved in a generic way from elements of differential geometry \cite{Flanders63}\cite{Choquet77}\cite{Eguchi80} considering that the contour of an n-chain $C_n$ is a $(n-1)$-chain, $C_{n-1}$ and that there is an operator $\partial$ that maps $C_n$ into $C_{n-1}$

\begin{equation}\label{opchain}
C_n \xrightarrow{\partial} C_{n-1} \Rightarrow \partial C_n = C_{n-1}.
\end{equation}

Even for closed chains that are called cycles, $Z_n$ we have

\begin{equation}\label{cycle}
\partial Z_n = 0,
\end{equation}
and which is a condition that can be applied to the Eq.(\ref{coint}). On the other hand there are some chains, $B_n$ which are contours of high dimensional chains

\begin{equation}\label{formB}
B_n = \partial C_{n+1}.
\end{equation}

In this sense we can write

\begin{equation}\label{opefb}
\partial B_n = 0,
\end{equation}
which therefore leads to

\begin{equation}\label{opec}
\partial (\partial C_{n+1}) = \partial^2 C_{n+1} = 0
\end{equation}

\subsection{Differential forms}

The ideas mentioned above can be expressed considering that the integral over a form $\omega_n$ is a scalar

\begin{equation}\label{cifoc}
\int_{C_n} \equiv \int_{C_n} f_{i_1...i_n}dx_{i_1} \wedge dx_{i_2} \wedge... \wedge dx_{i_n} = const.,
\end{equation}
where $ \wedge$ denotes wedge product, so that one can define an operator $d$

\begin{equation}\label{ped}
d \omega_n = \omega_{n+1}.
\end{equation}
and soon

\begin{equation}\label{oped}
d (d\omega_n) = d^2 \omega_n = 0.
\end{equation}

Eq.(\ref{oped}) makes an allusion to the well-known Poincare's lemma:

\emph{An exact $n$-form $\omega_n$ is the derivative of an $(n-1)$-form}
\begin{equation}\label{inopef}
\omega_n = d\omega_{n-1}.
\end{equation}
Thus we have for a class of forms

\begin{equation}\label{opef}
d(d\omega_{n-1}) = d^2 \omega_{n-1} = 0.
\end{equation}

Eq.(\ref{opef}) has a consequence from Stokes' theorem which states that if $\omega$ is a 
$p$-form and $C$ is a $p+1$-chain, that is, a contour of the $p$-form, then we can write that

\begin{equation}\label{ingefo}
\int_{\partial C} \omega = \int_{C} d\omega,
\end{equation}
which leads to Gauss and Stokes' theorem. In this sense we can now define Field Strength geometrically

\begin{equation}\label{fisgen}
F = \frac{1}{2}F_{\mu\nu}dx^{\mu}\wedge dx^{\nu}, 
\end{equation}
so that the equations of motion (Maxwell's equations) can be written as

\begin{equation}\label{opegfs}
d F = 0
\end{equation}
which allows us to geometrically generalize the Eq.(\ref{flength}) and 
Eq.(\ref{flestrength}) under these conditions. It is worth mentioning that this analysis can also be extended to non-Abelian systems in which fields such as Yang-Mills fields or analogues of Gauge fields are present 

\begin{equation}\label{ym}
W_{\mu\nu} = \partial_{\mu}W_{\nu} - \partial_{\nu}W_{\nu} + gW_{\nu}\times W_{\nu},
\end{equation}
with $W_{\mu}$ being a Gauge field, so that

\begin{equation}\label{opeym}
D^{\nu}W_{\nu\mu} = 0 \rightarrow \partial^{\nu}W_{\mu\nu} = -gW^{\nu}\times W_{\mu\nu},
\end{equation}
can still be included in geometric terms that satisfy the Eq.(\ref{coint}). 
As a conclusion from the discussion above, such elements of differential geometry lead us to think that the SSR could be a non-trivial space whose boundary region is a Weyl manifold integrable in the weak field limit. Therefore, SSR would be a non-trivial topological structure whose implications could be explored more deeply in a future work.

\section{\label{sec:level1} Cosmological implications of SSR: the equation of state (EOS) 
of vacuum} 

\subsection{Energy-momentum tensor in SSR} 

Let us write the $4$-velocity in the presence of 
$S_V$, as follows:
  
\begin{equation}
 U^{\mu}=\left[\frac{\sqrt{1-\frac{V^2}{v^2}}}{\sqrt{1-\frac{v^2}{c^2}}}~ , ~
\frac{v_{\alpha}\sqrt{1-\frac{V^2}{v^2}}}{c\sqrt{1-\frac{v^2}{c^2}}}\right],
\end{equation}
where $\mu=0,1,2,3$ and $\alpha=1,2,3$. The $4$-velocity of SR is naturally recovered if $V\rightarrow 0$. 

The energy-momentum tensor for a perfect fluid is 
given as follows: 

\begin{equation}
T^{\mu\nu}=(p+\epsilon)U^{\mu}U^{\nu} - pg^{\mu\nu},
\end{equation}
where $U^{\mu}$ is given in Eq.(61), $p$ is the pressure and $\epsilon$ is the energy density.

From Eq.(61) and Eq.(62), by calculating the component $T^{00}$, we get 

\begin{equation}
T^{00}=\frac{\epsilon(1-\frac{V^2}{v^2})+p(\frac{v^2}{c^2}-\frac{V^2}{v^2})}{(1-\frac{v^2}{c^2})}
\end{equation}

From Eq.(63), if we take the limit $V\rightarrow 0$, we obtain the well-known component $T^{00}$ 
of Relativity. 

\subsection{The vacuum limit $S_V$ and the EOS of
vacuum}

Now, in order to obtain $T^{00}$ in Eq.(63) for the vacuum limit given at the ultra-referential-$S_V$, i.e., in the absence of matter, we perform the calculation, namely: 

\begin{equation}
 lim_{v\rightarrow V} T^{00}= T^{00}_{vacuum}=\frac{p[(V^2/c^2)-1]}{[1-(V^2/c^2)]}= -p 
 \end{equation}

 According to Eq.(64), as we must have $T^{00}>0$, i.e., always a positive energy density, we get $p<0$, which implies in a negative pressure for the vacuum energy density of the ultra-referential-$S_V$. Thus, we verify that a negative pressure emerges naturally from the energy-momentum tensor of SSR in the limit of $S_V$
 ($v\rightarrow V$). 

So we can obtain $T^{\mu\nu}_{vacuum}$ by calculating the following limit: 
 
 \begin{equation}
 T^{\mu\nu}_{vacuum}= lim_{v\rightarrow V}T^{\mu\nu}= -pg^{\mu\nu},
 \end{equation}
 where we naturally conclude that the vacuum 
 energy density is $\rho_{\Lambda}=\epsilon=-p$.
 
 We realize that $T^{\mu\nu}_{vac.}$ is a diagonalized tensor. Thus, the vacuum-$S_V$ 
 in the space-time of SSR is a fluid in equilibrium with a negative pressure. This leads to a cosmological anti-gravity represented by the 
 cosmological constant $\Lambda$ and thus the accelerated expansion of the universe. 

\section{\label{sec:level1} Conclusions and prospects}      

First of all, based on the quantum principle of zero-point energy that originates from the uncertainty principle, which is not consistent with the classical space-time of Special Relativity (SR), we have searched for a quantum space-time structure to be consistent with the idea of the absence of rest in the quantum world by precisely postulating an invariant minimum speed $V$, which allowed us to include the concept of vacuum associated with a preferred reference frame $S_V$ (Fig.1). 

The minimum speed is a new kinematic invariant given for lower energies, which led to a new Deformed Special Relativity (DSR) so-called Symmetrical Special Relativity (SSR), from where there emerged the cosmological constant, which allowed us to show the equivalence of SSR-metric with a dS-metric. The small order of magnitude of the cosmological constant ($\Lambda\sim 10^{-35}s^{-2}$) 
was estimated successfully. 

Finally, we were able to obtain the tiny numerical value of the cosmological constant $\Lambda_0=1.934\times 10^{-35}s^{-2}$ given in the zero-gravity limit (flat universe) when considering the redshift $z=1$ and the Hubble time $\tau_0$ at which the universe goes over from a decelerating to an accelerating expansion. 

Section 5 was dedicated to the Weyl geometrical structure of SSR. We have shown that the factor $\Theta(v)$ in Eq.(3) behaves like a conformal Weyl factor, so that SSR includes a Weyl conformal geometry in the regime of Newtonian weak-field given in the SSR scenario 
($V<<v<<c$). Such regime corresponds to a slight acceleration of the universe given for redshift $z=1$, where we have obtained the tiny value of the cosmological constant according to the experiments. So we have concluded that the current expanding universe is governed by a Weyl conformal geometry for weak-field ($\phi/c^2<<1$) by representing an almost flat space-time as a special case of Eq.(3) of SSR. 

Here it is important to call attention to a chapter of a book that 
was the first publication that considered the connection between the minimum speed and the Weyl geometry\cite{Weyl1}. 

The investigation of symmetries of the SSR-theory should be done by means of their association with a new kind of electromagnetism when we are in the limit $v\rightarrow V$, which could explain the problem of high magnetic fields in magnetars\cite{gravastar}\cite{gravastar1}\cite{Greiner}, super-fluids in the interior of gravastars\cite{gravastar1} and other types of black hole mimickers\cite{haw}\cite{Tds,Tds1,haw2}.

SSR theory has strong implications for quantum field theories (QFT) and Feynman's rules, since the dispersion relation of special relativity ($E^2=p^2-m^2$, with $c=1$)
is modified in the presence of the minimum speed, i.e., the new dispersion relation is 
$E^2=p^2-m^2(1-V^2/v^2)$ or $E^2=p^2- m^2+(m^2V^2/v^2)$.

We should note that the SSR dispersion relation has an additional mass term $m^2V^2/v^2$. If $v=V$, such a term would nullify the mass term $m^2$, and then we would have a dispersion relation for a massless particle as if it were a photon.

The additional term in the dispersion relation suggests that the minimum speed has a deep implication on QFT propagators, as for instance, the known propagator of a free electron to be renormalized, i.e., $1/(p^2-m^2\pm i\epsilon$), where $\pm i\epsilon$ comes from 
calculating residuals with the aim of renormalizing the integral. We must realize that the new propagator in the integral presents the new SSR dispersion relation, i.e., $1/[p^ 2-m^2+(m^2V^2/v^2)]$. Now we see that the new term $m^2V^2/v^2$ seems to provide a physical basis for the renormalization, since it seems to play the role of the mathematical term $\pm i\epsilon$, i.e., $\pm i\epsilon\equiv m^2V^2/v^2$. If this is true, then the minimum speed would be a kind of natural cut-off for the infinities that appear in quantum field theories (QFT), which could also provide a fundamental explanation for Feynman's rules. This vast issue will be explored in future works and it actually seems to be very promising. 

Quantum Electrodynamics (QED) is a very successful theory,i.e.,although infinities appear, the theory is renormalizable, which means that infinities can be removed using various techniques. One of the great successes of this theory is the precision with which it obtains the Lambda shift between two energy levels in the Hydrogen atom. Electromagnetic interactions between the proton and the electron, where the fine structure constant 
($e^2/\hbar c$) plays a fundamental role, are responsible for the Lambda shift. In view of this, it would be interesting to investigate how the minimum speed will interfere with the Lambda shift. At first sight, we can already realize that the presence of the minimum speed will interfere so little with the Lambda shift, since the minimum speed has a gravitational origin, i.e., $V=\sqrt{Gm_pm_e} e/\hbar$\cite{N2016}, so that the gravitational interaction between the proton and the electron, which is proportional to $Gm_pm_e\propto V^2$ is responsible for an almost negligible perturbation in the Lambda shift, which is due to the fact that the gravitational interaction is much weaker than the electromagnetic interaction. This subject will be deeply investigated in future work.

As the SSR theory has a kinematic basis, it can be applied in Relativistic Hydrodynamics with the purpose of investigating superfluids in which the existence of a privileged reference frame plays an important role in these models, as was recently shown by 
Santos et al.\cite{Santos}. 

In sum, the perspectives opened by SSR suggest that an invariant minimum speed in nature produces a set of kinematic transformations that provide the basis for a series of models of scalar fields in the cosmological scenario by allowing to obtain the numerical value of the cosmological constant as well as the exploration of models of compact astrophysics systems.

\end{document}